\documentclass[10pt]{iopart}
\usepackage{graphicx}
\def\h{{\bf h}}
\def\F{{\bf F}}
\def\bfk{{\bf k}}
\def\p{{\bf p}}
\def\k{{\bf k}}
\def\q{{\bf q}}
\def\gs{g_{\rm s}}
\def\alphas{\alpha_{\rm s}}
\def\alphaEM{\alpha_{\rm EM}}
\def\mD{m_{\rm D}}
\def\nf{N_{\rm f}}

\def\lsim{\mbox{~{\protect\raisebox{0.4ex}{$<$}}\hspace{-1.1em}
	{\protect\raisebox{-0.6ex}{$\sim$}}~}}
\begin{document}

\title[Electromagnetic Emission and Energy Loss in the
QGP]{Electromagnetic Emission and Energy Loss in the QGP}

\author{Guy D.\ Moore\dag }

\address{\dag\ Physics Department, McGill University, 3600 University
	Street, Montreal QC H3A 2T8, Canada }

\ead{guymoore@physics.mcgill.ca}

\begin{abstract}
I discuss why photon production from the Quark Gluon Plasma (QGP)
presents an interesting problem, both experimentally and theoretically.
I show how the photon emission rate can be computed under the
simplifying assumption that the QGP fully thermalizes.  The theoretical
issues are very similar to those for jet energy loss; so it should be
possible to treat them in a common formalism and relate the predictions
of one phenomenon to those of the other.
\end{abstract}

\submitto{\JPG}

\section{Photons as a deep probe}

The main goals of the RHIC experimental program are, to produce and to
characterize the quark-gluon plasma (QGP).  The plasma is defined as being a
state of matter for which the density of partons (quarks and gluons) is
so great that a description in terms of hadronic degrees of freedom is
impossible, which is furthermore extensive and relatively close to
equilibrium.  The problem is that the QGP expands, cools, and
hadronizes.  Even after hadronization, hadrons
continue to interact; much of the information about
the initial state is therefore lost.  For instance, the spectrum of
hadronic final states which we observe at RHIC is well characterized by
a thermal model with a temperature of only about 160 MeV, too low for a
QGP but reasonable for the freezeout of a hadronic gas
\cite{thermal_fits}.  Therefore, the hadronic observables give only
rather indirect information about the QGP phase of the collision.

For this reason, it was proposed years ago to study ``hard'' probes,
meaning particles which are produced early in the collision, deep
within the QGP, and which then escape with little or no interaction,
giving relatively direct information about the early stages of the QGP.
Shuryak proposed photons as a promising direct probe
\cite{Shuryak}, and they continue to be actively investigated
\cite{yellow_book}.

The advantage of photons as a probe of the QGP is that any photons
produced are almost sure to escape without re-interacting, so they give
direct, rather than indirect or processed, information about the early
stages of the heavy ion collision.  The main disadvantage is that there
are several sources of photons, with the concomitant problem of
determining which photons arose from which source.  Namely, there are
\begin{itemize}
\item
Prompt photons, those produced by the initial collisions of
the partons which constitute the heavy ions being collided;
\item
QGP photons, those produced during scatterings and
interactions within the quark gluon plasma;
\item
Hadronic photons, those produced after the QGP has changed
into a hadron gas but before it has spread out enough that mutual
interactions cease; and
\item
Decay photons, those produced in the decay of certain hadrons,
chiefly $\pi^0$ and $\eta$ mesons, ``long'' after the collision has
finished. 
\end{itemize}

\begin{figure}[h]
\centerline{
\includegraphics[width=1.5in,height=1.9in]{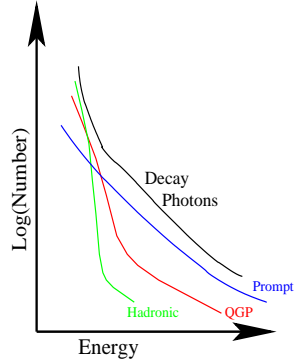}}
\caption{\label{fig1} Cartoon of the spectrum of each kind of photon
produced in a heavy ion collision.  Decay photons are expected to
dominate at essentially every energy.}
\end{figure}

Of the ``direct'' photons (those produced in the collision rather than
in much later decays), the QGP and hadronic photons are expected to
dominate at lower energies, and the prompt photons at high energies.
This is because there are more collisions during the QGP phase than
between the initial state partons, so there are more
opportunities to produce
photons; but the energies of the particles in the QGP are smaller than
the energies of the primary hadrons, so the produced photons are softer.

The problem is that decay photons are expected to dominate over direct
photons at essentially every energy (see Fig.\ \ref{fig1}), and they
carry no information about the early stages of the collision.  Arguably
the most interesting photons are the QGP photons; the prompt photons are
also interesting, as they give us information about the initial parton
content of the nuclei and therefore improve our normalization for jet
quenching and other hard probes.

One way to understand the dominance of the decay photons is to
note that the fine structure constant $\alphaEM$ is small,
and to ask how the number of each photon type scales with $\alpha_{\rm
EM}$.  The prompt, QGP, and hadronic photon numbers all scale as
$\alphaEM$, because the duration of the phases they emerge from,
and the energies of the particles involved, are roughly independent of
$\alphaEM$, whereas the processes that produce them (obviously)
involve electromagnetic interactions.
The number of decay photons, on the other hand, is
nearly independent of $\alphaEM$; if $\alphaEM$ were
smaller, the $\pi^0$ mesons would be longer lived, but would still
eventually decay into photons, so the number of produced photons would
be unchanged (modulo the few photons from $\eta$ decay).

At this meeting, we heard that prompt photons have now been convincingly
detected by the PHENIX collaboration \cite{PHENIX_gamma}.
\begin{figure}[h]
\vspace{1.4in}
\includegraphics{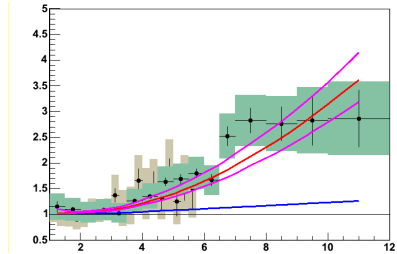}
\caption{\label{fig2} PHENIX measurements of photon excess over expected
background from $\pi^0$ decay, in central $Pb{-}Pb$ collisions.  (The
red curve is the theoretical expectation.)  A positive signal is seen
from 4 GeV on up.}
\end{figure}
\noindent
Fig.\ \ref{fig2} shows that the detection only extends from about 4 GeV
up, which as we will see means that only prompt photons have been
unambiguously observed (so far).  The WA98 collaboration at the SPS has
claimed an observation of QGP photons as well \cite{WA98}, but the
statistical significance is not that strong.  We expect the PHENIX
experimental results to improve, certainly tightening our determination
of the prompt flux, but also hopefully detecting the QGP photons.  This,
together with a theoretical analysis of the QGP photon flux, will give
us a direct handle on the early stages of heavy ion collisions.

\section{Computing the photon emission rate}

Computing the photon production by a QGP turns out to be an interesting
theoretical problem.  Very little is known or understood about the very
early stages of the QGP, before it approximately thermalizes, so we will
treat only the simpler problem of photon emission from a thermalized
QGP, in the approximation that the expansion time of the QGP is long
compared to the dynamical timescales associated with the photon
production.  We will furthermore make an expansion in $\alphas
\ll 1$, even though it is not obviously justified.  This turns out to
make the slow expansion rate a consequence of the assumption of
near-equilibrium. 

\begin{figure}[t]
\centerline{
\includegraphics[width=2in,height=0.9in]{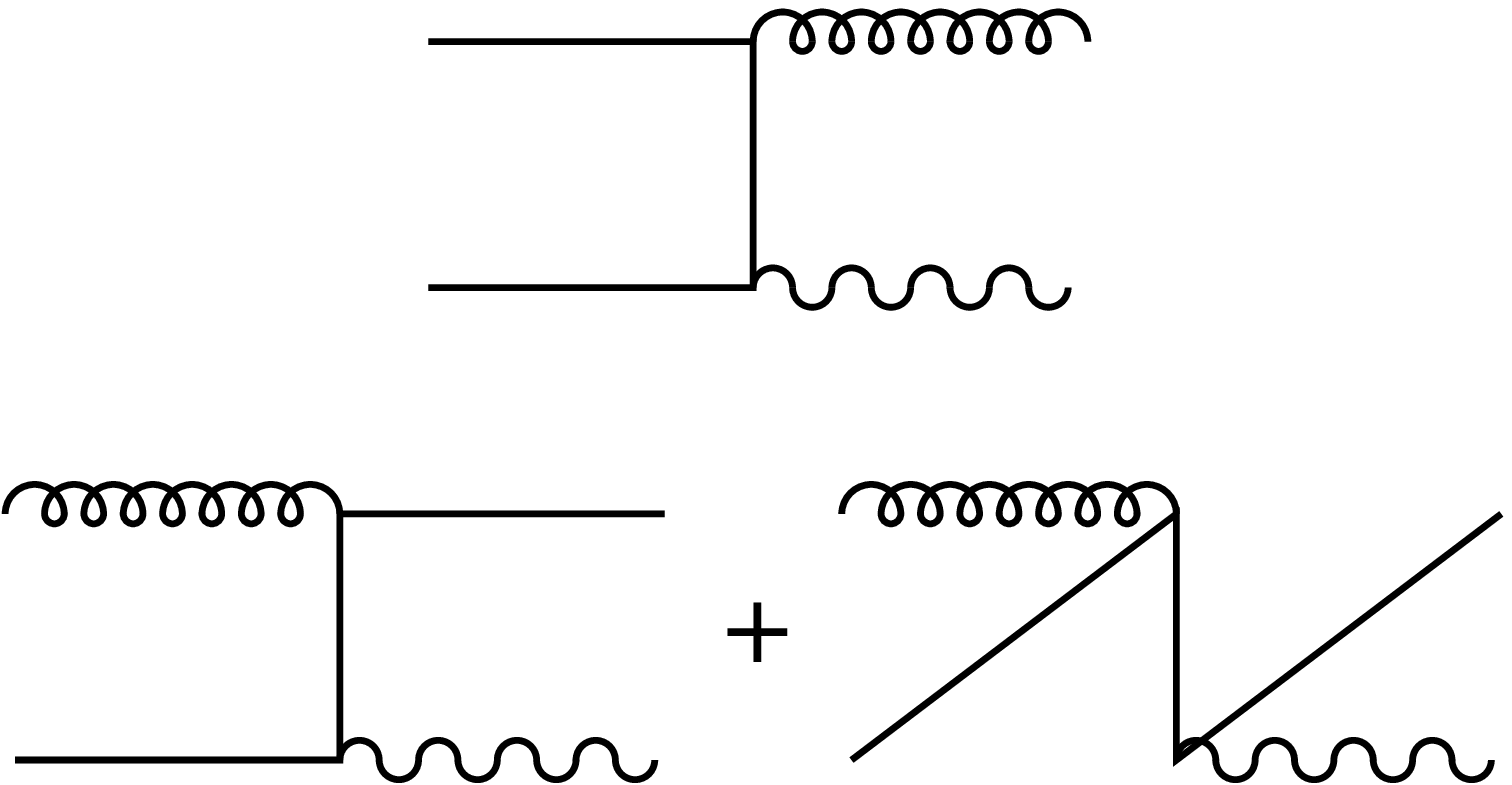}
\hspace{0.4in}
\includegraphics[width=2.2in,height=0.8in]{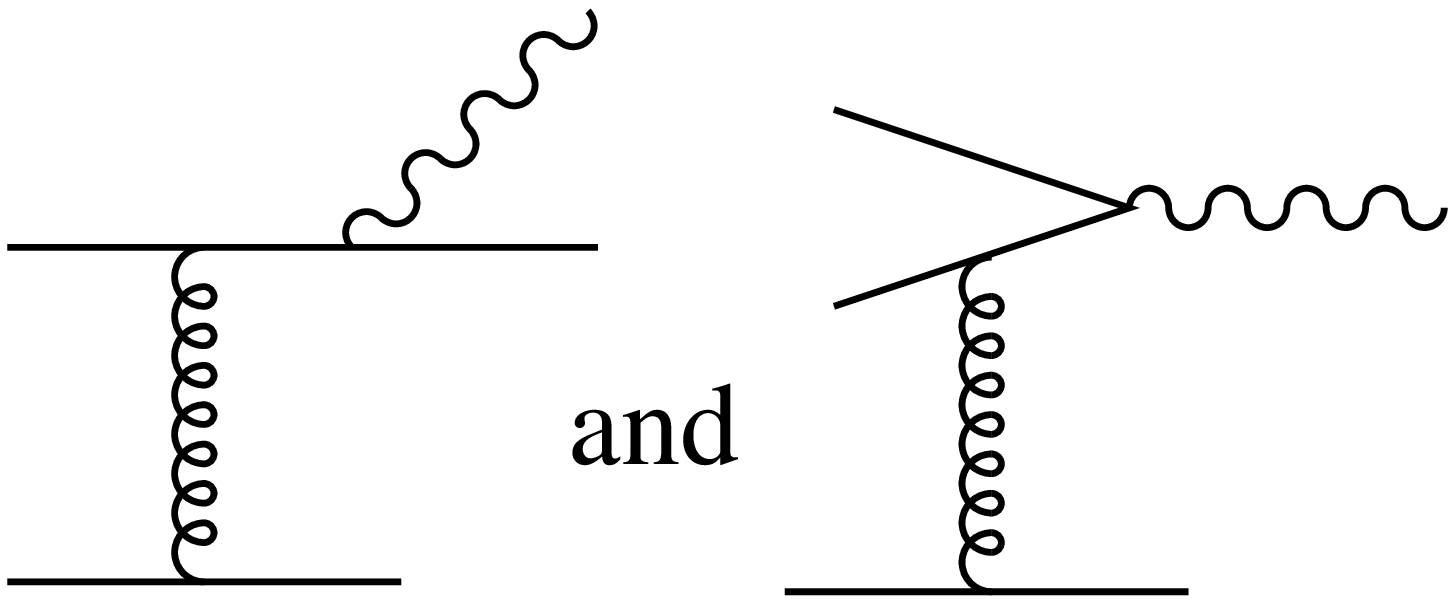}}
\caption{\label{fig3} Left:  naively leading order processes for photon
production.  Right:  processes niavely higher order, but contributing at
the same order.}
\end{figure}

This problem was thought solved in 1991, when two groups
\cite{Seibert,Baier} evaluated the leading order diagrams for the
process, shown on the left in Fig.\ \ref{fig3}.  This is not quite as easy as it
looks, because the intermediate propagator receives plasma corrections,
described by Braaten and Pisarski's Hard Thermal Loops \cite{Braaten}.
Their result was,
\begin{equation}
\hspace{-1.0in}
\frac{d\Gamma}{dk} = 
\frac{2\alphas \alphaEM k T^2}{\pi [e^{k/T}+1]} \:
	\left[ \ln \left( \frac{3k}{2\pi \alphas T} \right)
	-\frac{1}{2} - \gamma_E + \frac{4 \ln 2}{3} 
	+ \frac{\zeta '(2)}{\zeta(2)} + O(T/k) 
	\right] \left[ \sum_q Q^2_q
	\right].
\label{naive_leading}
\end{equation}
Here $\sum_q$ runs over all light ($m \ll T$) quarks, and 
$Q_q$ is the electric charge of the quark; so the sum is $5/9$ if the
strange quark is taken to be heavy and $2/3$ if it is taken to be
light.  Note that the result is logarithmically enhanced above the naive
$O(\alphas \alphaEM)$ expected from vertex counting.  This
occurs from the region where a spacelike propagator becomes soft.

This treatment is incomplete, however, because it misses
contributions which are naively suppressed by a further power of
$\alphas$, but which receive collinear enhancements which render
them also $O(\alphas \alphaEM)$.  This was first noticed
by Aurenche, Gelis, Kobes, and Zaraket \cite{Aurenche_etal}, who were
computing the next order corrections to Eq.~(\ref{naive_leading}),
arising from the diagrams on the right of Fig.~\ref{fig3}, and
found that they were of the same order.

These diagrams correspond to bremsstrahlung and processes related by
crossing symmetry, and they encounter the same collinear enhancements as
bremsstrahlung usually encounters.  The cross-section for soft
$t$-channel scattering is power (Coulomb) divergent in vacuum; this is
cut off by plasma effects in the QGP medium, leading to a cross-section
which is $O(\alphas)$, rather than the naive $O(\alpha_{\rm
s}^2)$.  Hard bremsstrahlung of a photon is suppressed by a further
power of $\alphaEM$; even though the scattering is soft and the
photon is hard, there is no additional suppression if the photon is
emitted collinear to the emitting particle.

One way of understanding this collinear enhancement is to
think about the spacetime picture of how the emission occurs.
\begin{figure}[h]
\centerline{
\includegraphics[width=2.2in,height=1.0in]{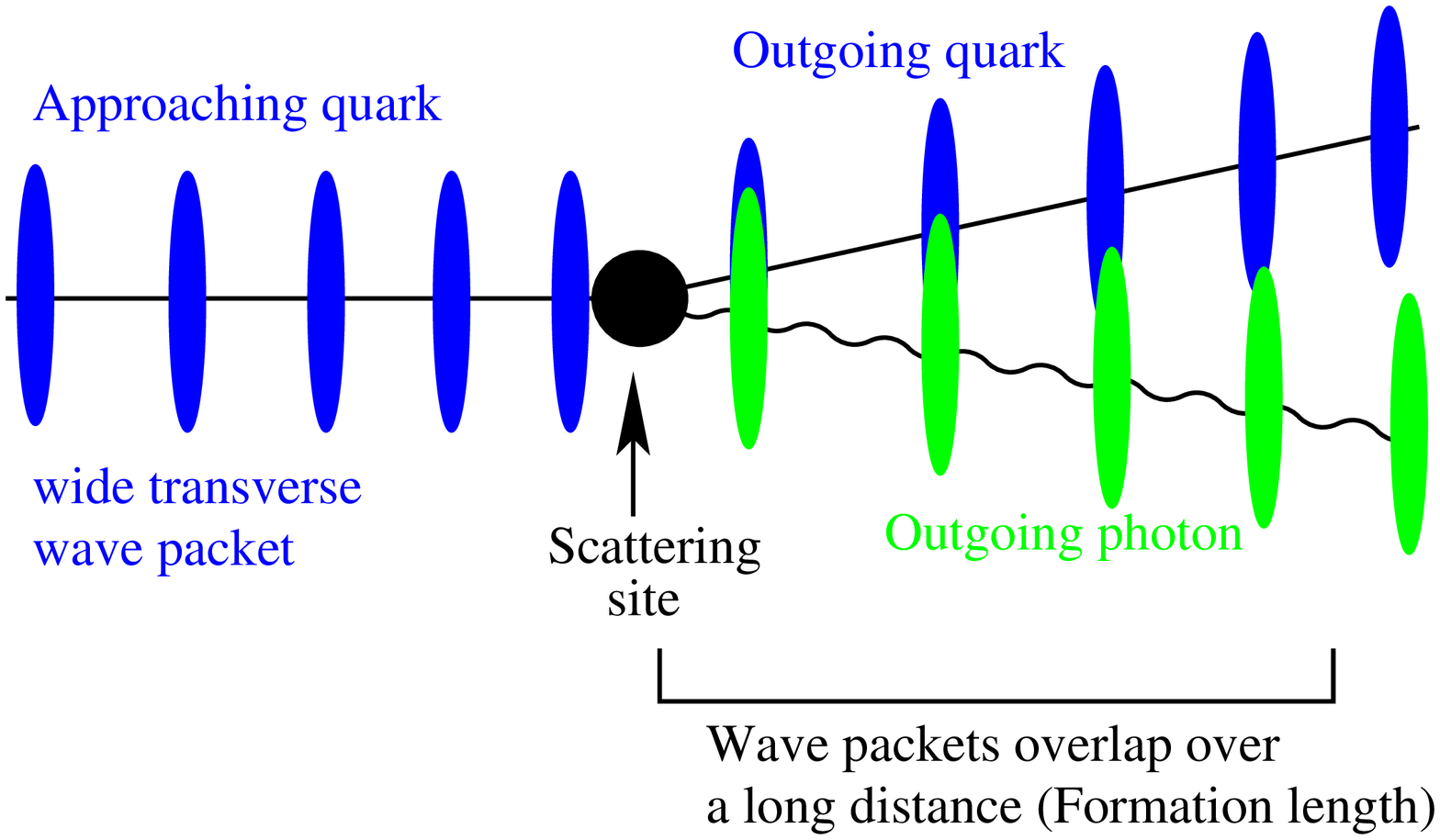}
\hspace{0.3in}
\includegraphics[width=2.2in,height=0.7in]{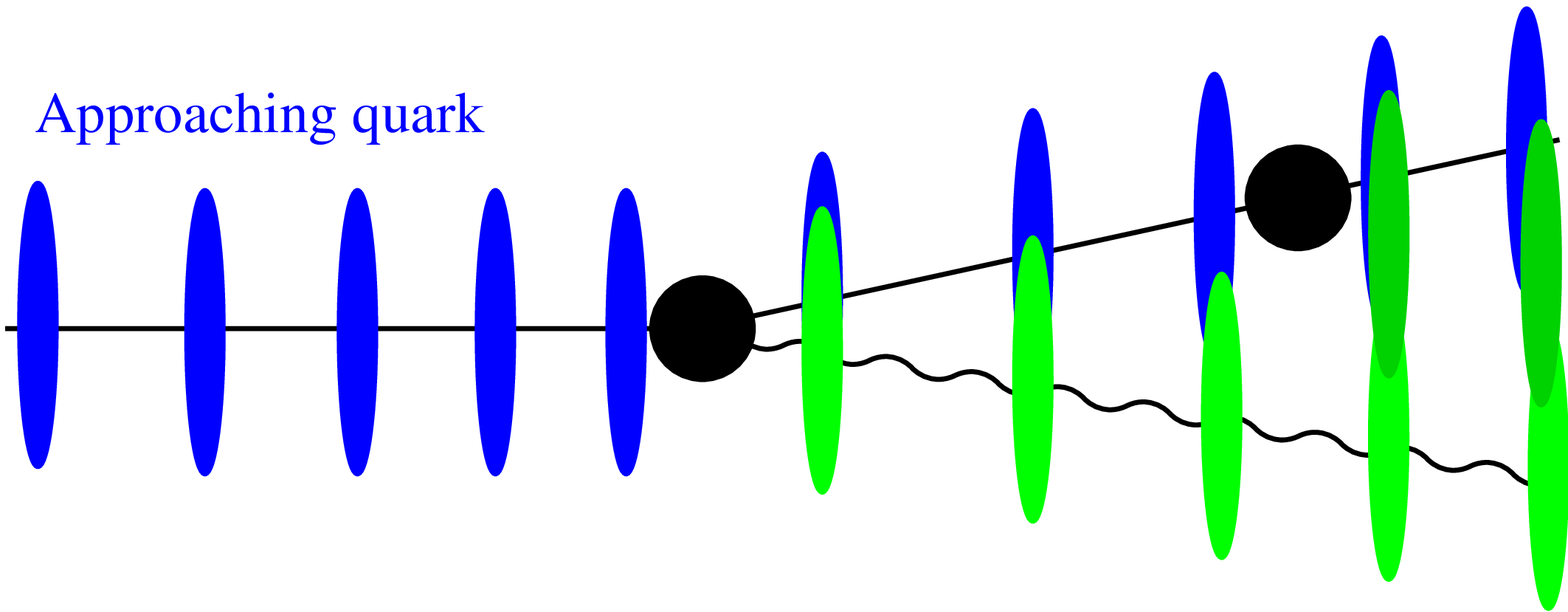}}
\caption{Left:  spacetime picture of photon emission.  Right:
interference of photon emissions from sequential scatterings. 
\label{fig5}}
\end{figure}
For small angle scattering ($\sim g_{\rm s}$) and a small photon
opening angle ($\sim g_{\rm s}$), one allow the particles 
wave packets with a large transverse size ($\sim 1/g_{\rm s} T$).
The combination of small
opening angle and large transverse wave packet means that the photon and
quark wave packets overlap for a long time ($\sim 1/\alphas T$),
leading to a large coherent enhancement of the emission.  The time
during which the wave packets overlap is called the coherence time.
Note that it is essential to this argument that the quark move close to
the speed of light; heavy quarks to not bremsstrahlung radiate
efficiently.

This spacetime picture also shows why treating only these additional
diagrams is not sufficient.  An additional scattering before the
coherence time of the emission will lead to a second photon emission
amplitude which overlaps and will interfere with the first:
This will happen $O(1)$ of the time, because the mean path between
scatterings, $O(1/\alphas T)$, coincides with the mean formation
time, for thermal photon production.  [For production of very soft
bremsstrahlung or very hard pair annihilation photons, the interference
effects are parametrically large.]  A lengthy power counting \cite{AMY2}
shows that diagrams with repeated gluon ``rungs,'' 
\begin{figure}[h]
\centerline{
\includegraphics[width=2in,height=0.54in]{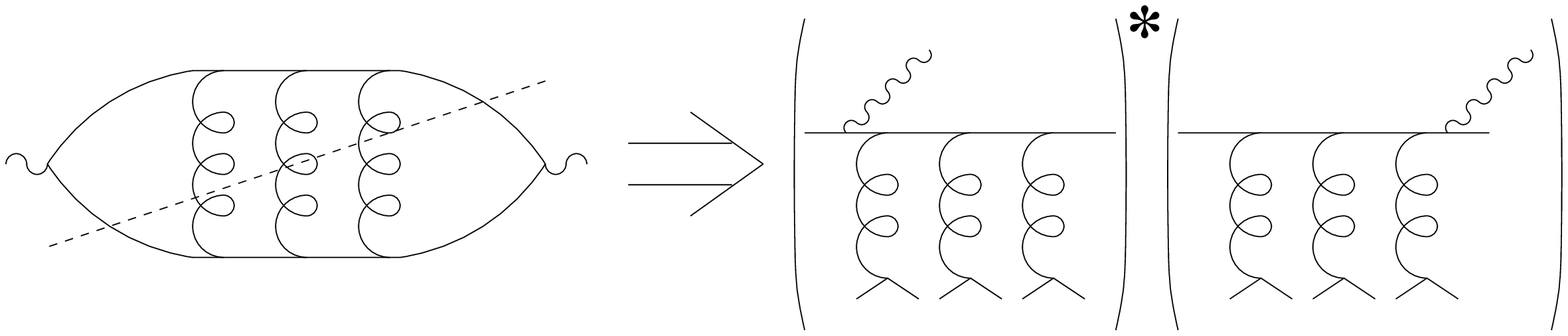}}
\caption{Diagrams, and interference effects, requiring resummation
	\label{fig7}}
\end{figure}
corresponding to interference before and after $n$ scattering processes,
must be resummed--but that no other processes contribute at leading
order in $\alphas$.  The interference between photon emissions
associated with successive scattering events is destructive, reducing
the number of photon emissions from what we would find if each
scattering event produced photons independently.  This interference
effect was first noticed, in a purely QED setting, by Landau,
Pomeranchuk, and Migdal \cite{LPM}, and is called the {\sl LPM effect}.
In the QCD context, it has been considered by a few groups
starting in the mid-1990's \cite{BDMPS,Zakharov}.

Sufficiently infrared modes in the QCD plasma, with wave number $k \lsim
\alphas T$ (the inverse magnetic screening length), are beyond
analytic treatment even at weak coupling.  Fortunately, the analysis
\cite{AMY2} shows that photon emission is not sensitive to this scale at
leading order.  [Note that the entire
treatment we are discussing is framed upon the assumption that a formal
expansion in $\alphas \ll 1$ is justified; we keep all effects
which are formally leading in such an expansion, and make all
permissible approximations which are valid at leading order.  It is not
obvious whether the expansion in $\alphas$ is even qualitatively
reliable for the environment actually encountered in the QGP, and it
will be difficult even to address that question until we are able to
perform a calculation beyond leading order.]

Resumming the diagrams for photon emission with multiple scatterings
leads to a Boltzmann-like equation, which describes the evolution of the
density matrix $| {\bf p}_q (t) \rangle \langle ({\bf p}-{\bf
k})_q,k_\gamma (t)|$, representing the interference between the process
where a photon has and has not been emitted, as the quark undergoes
scatterings in the medium.  The photon emission rate is,
\begin{eqnarray}
\label{eq:brem1}
    \frac{dN_\gamma}{d^3 \bfk d^4 x} 
    & = &
    \frac{2 \alphaEM}{4\pi^2 k}
    \int_{-\infty}^{\infty} \frac{dp}{2\pi}
    \int  \frac{d^2 \p_\perp}{(2\pi)^2} \;
	\frac{n_f(k{+}p) \, [1{-}n_f(p)]}
	{2[ p \, (p{+}k) ]^2} \times \nonumber \\ && \qquad \times
	    \left[ p^2 + (p{+}k)^2 \right]
    {\rm Re} \, \Big\{ 2 \p_\perp \cdot{\bf f}(\p_\perp;p,k) \Big\} \\
\label{eq:brem2}
2\p_\perp & = & i \delta E \; {\bf f}(\p_\perp;p,k) +
	{2\pi \over 3} \gs^2 \int
	\frac{d^2 q_\perp}{(2\pi)^2}
	\frac{\mD^2\, T}{q_\perp^2(\mD^2{+}q_\perp^2)}\times
\nonumber \\ && \qquad \times
	\left[ \, \strut
	    {\bf f}(\p_\perp;p,k) - {\bf f}(\q{+}\p_\perp;p,k)
	\right]^{\strut} , \\
\label{eq:brem3}
\delta E & = &
     \frac{\p_\perp^2 + m_\infty^2}{2} \;\,
	\frac{k}{p ( k{+} p)} \, .
\end{eqnarray}
The first expression relates the rate of photon emission to the solution
to an integral equation, describing the propagation of the quark or
quark plus photon system through the plasma.  The second equation is the
integral equation; the first term is a ``dephasing'' term recording how
the photon and quark wave packets evolve off from on top of each other,
and the second term is a collision term, reminiscent of the collision
term in a Boltzmann equation.  The third equation expresses $\delta E$,
the energy difference between the system with a quark of momentum $\p$
and a quark plus photon of momentum $(\p{-}\k),\k$; this energy
difference is responsible for the ``dephasing.''
Here $\p_\perp$ is the component of $\p$ perpendicular to $\k$, and
$m_{\infty}^2 = \gs^2 T^2/3$ is the thermal ``mass'' of a quark,
which parametrizes how forward scattering in the medium changes the
dispersion relation of the quark.
This is relevant because, as we saw, the coherence length is smaller if
the quark moves slower than the speed of light.

\begin{figure}[t]
\centerline{
\includegraphics[width=2.3in,height=2.3in]{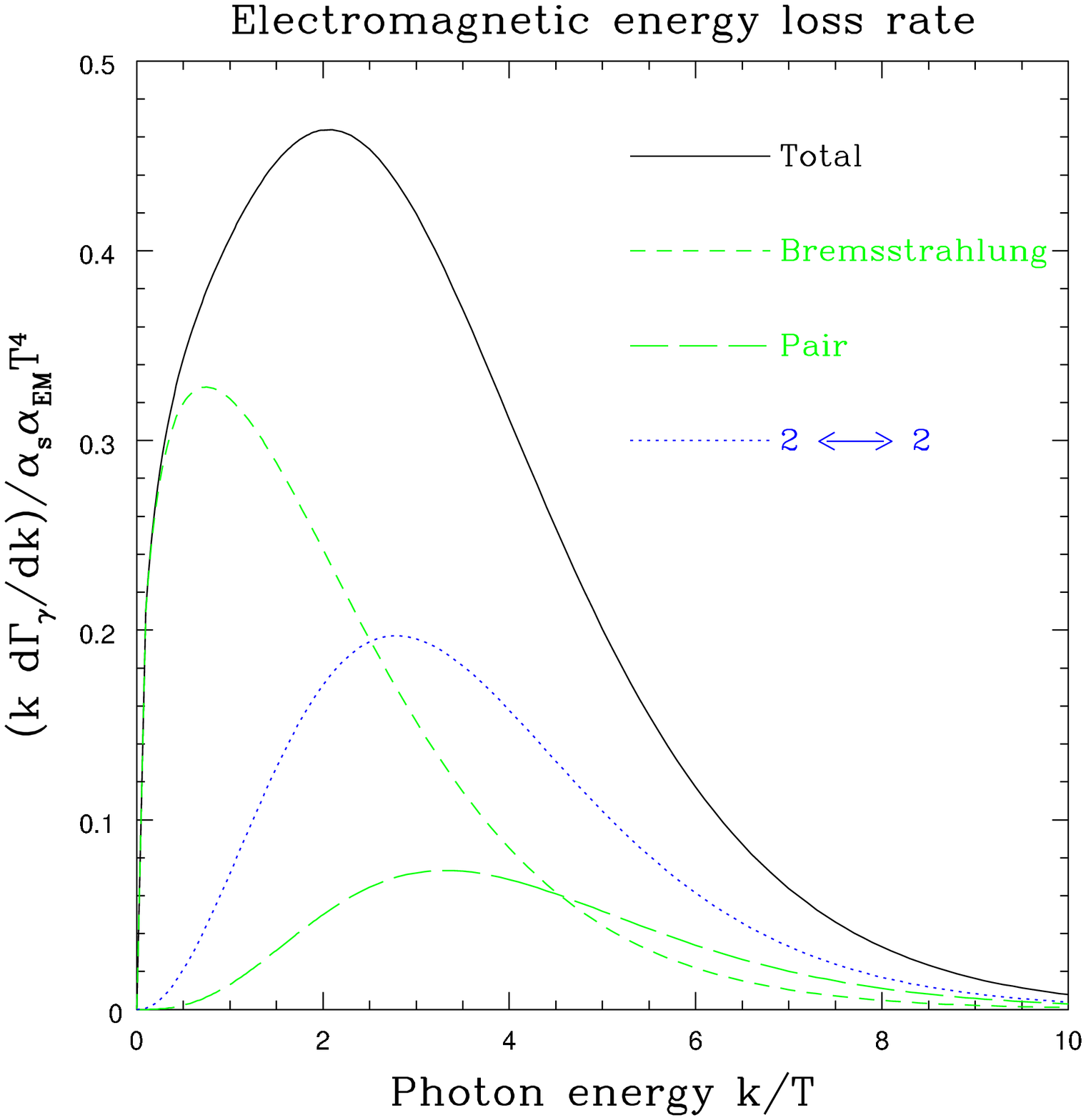}
\includegraphics[width=2.3in,height=2.3in]{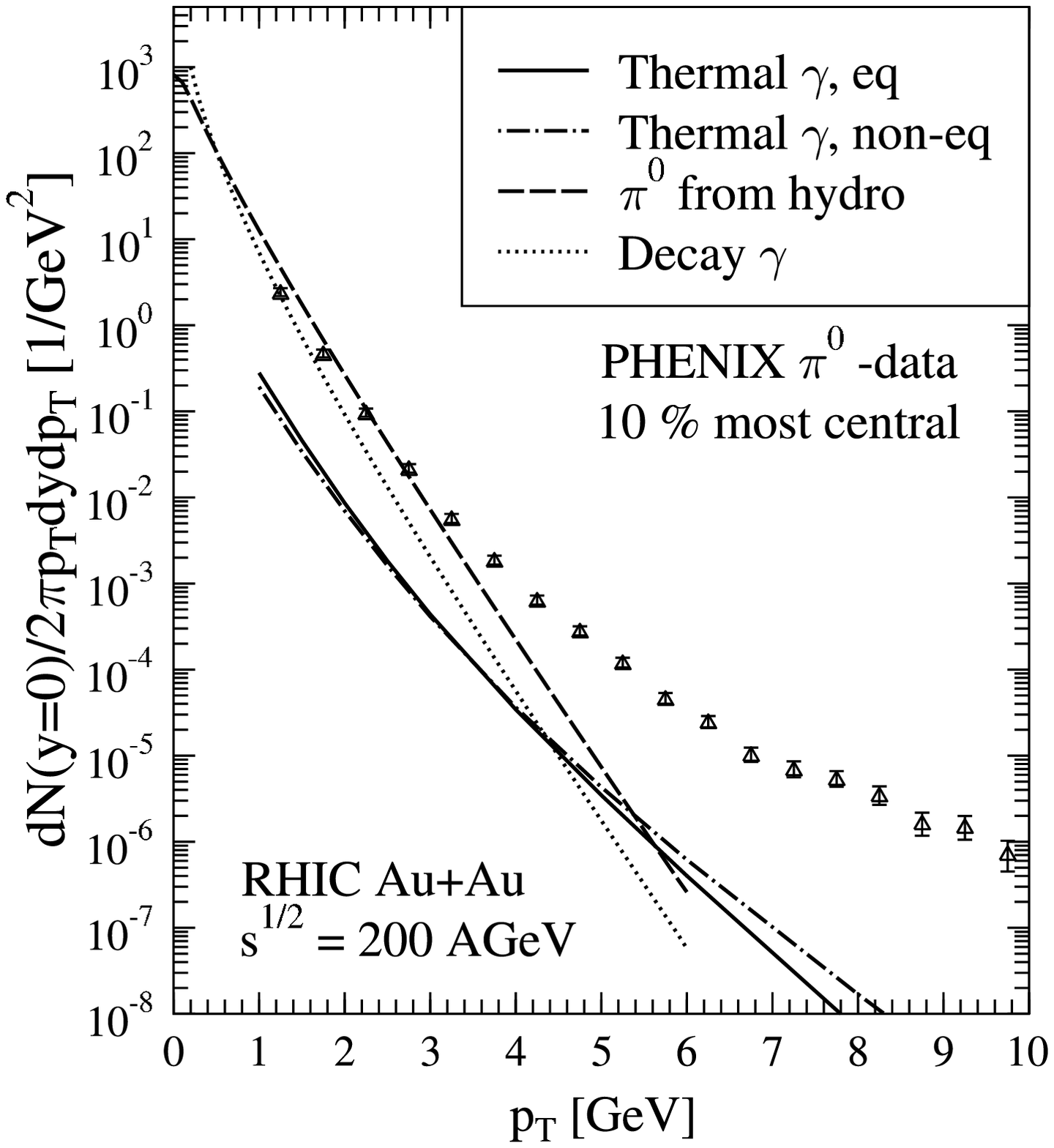}}
\caption{\label{fig8} Left:  Spectrum of photon energy produced from the
QGP, from each type of process.  Right:  hydro results of Ruuskanen {\it
et.\ al.\ } using these photon production rates.}
\end{figure}

This integral equation does not have an analytic solution
(that we know of).  It can be solved numerically by a
variational technique \cite{AMY3}, or by Fourier transforming to impact
parameter space and evolving an ODE \cite{AGMZ}.  A few parameter fit of
the result, is that the bracketed quantity in Eq.~(\ref{naive_leading})
should be increased by
\begin{equation}
      \sqrt{1 {+} {\textstyle {1\over6}} \nf}
      \left[
	\frac{1.096 \, \log(12.28 + 1/x)}{x^{3/2}}
	+ \frac{0.266 \, x}{\sqrt{1+x/16.27}} 
    \right] ,
\end{equation}
where $\nf$ is the number of light quark flavors (presumably 2 or 3).
The result is also displayed, for 2 flavors and $\alphas = 0.2$, in
Fig.\ \ref{fig8}.  Numerically, we find that the result only deviates
about $20\%$ from the answer we would obtain by ignoring the LPM effect
and treating the emission from each scattering to be independent
(which is equivalent to evaluating Eq.~(\ref{eq:brem2}) at linear order in the
collision term).  At very low and very high energies this statement
breaks down, but it is valid out past $10 T$ (say, 3 GeV).

How many photons are produced from the QGP?  What we have computed is
the production rate from a region of equilibrated QGP at some
temperature.  To get the total number from the QGP, this must be folded
over the hydrodynamical evolution of the plasma.  Our results for QGP
emission rates have been used by various hydro groups; for instance, the
results of Ruuskanen {\it et.\ al.\ } \cite{Rasanen} 
are shown in Fig.~(\ref{fig8}).
Generally, the uncertainties from the hydro evolution (even when
constrained to describe correctly the multiplicity and temperature of
the final state hadrons) are larger than the uncertainties from the
photon production rate \cite{Ruuskanen}.

The figure suggests that the photon production by QGP is about a
factor of 10 smaller than the background from $\pi^0$ decay; above
3 GeV, the $\pi^0$ rate dominates even more.  That means that the photon
production  (technically, the ratio of photons to $\pi^0$) must be
measured to better than $10\%$ at about 3 GeV, which should be quite
challenging but may be experimentally possible. 

\section{Jet quenching}

Actually, we know that the QGP will {\em not} be a fully thermal bath.
The particles which take the longest to thermalize are the highest
energy ones; therefore one expects a power law, high energy tail of
particles moving through the approximately thermalized QGP.  This high
energy tail is responsible for high energy jets, and their interaction
with the plasma, particularly their energy loss as they traverse the
plasma, is the source of jet quenching.  The dominant energy loss
mechanism is the bremsstrahlung of gluons--a process very similar to the
bremsstrahlung of photons, which we have just discussed.  This leads to
two conclusions:
\begin{enumerate}
\item
We should be able to study jet energy loss with the same formalism as
photon production;
\item
High energy particles moving through the QGP should also radiate
photons, which will give a power-law, nonthermal extra contribution to
the QGP photon production rate.  This can be computed in parallel with
jet energy loss.
\end{enumerate}
The issue of jet energy loss has been looked at by several authors and
has been nicely summarized in these proceedings by Ivan Vitev
\cite{Vitev}.  Nevertheless, we think it would be beneficial to treat it
again, being careful to work systematically to leading order in
$\alphas$, not treating the LPM effect as parametrically large
(especially as we find that it is usually quite small), and treating the
photon production in parallel.  This should give a concrete relation
between the extent of energy loss and the number of hard photons emitted
as hard partons traverse the QGP.

The rate at which a parton of momentum $p$ emits a gluon of momentum
$k$ in traversing the QGP, per unit time and $k$, is
\begin{eqnarray}
\label{eq:dGamma}
\hspace{-0.5in}
\frac{d\Gamma(p,k)}{dk dt} \hspace{-0.5in} & = & \frac{C_s \gs^2}{16\pi p^7} 
        \frac{1}{1 \pm e^{-k/T}} \frac{1}{1 \pm e^{-(p-k)/T}}
\left\{ \begin{array}{cc} 
        \frac{1+(1{-}x)^2}{x^3(1{-}x)^2} & q \rightarrow qg \\
        N_{\rm f} \frac{x^2+(1{-}x)^2}{x^2(1{-}x)^2} & g \rightarrow qq \\
        \frac{1+x^4+(1{-}x)^4}{x^3(1{-}x)^3} & g \rightarrow gg \\
        \end{array} \right\} \times \nonumber \\ && \times
\int \frac{d^2 \h}{(2\pi)^2} 2 \h \cdot {\rm Re}\: \F(\h,p,k) \, ,
\end{eqnarray}
with $C_s$ the relevant Casimir, $C_s = (4/3)$ except for $g \to gg$,
for which it is 3.  $\h$ is the non-collinearity, $\p \times \k$, and
$\F$ is the solution to
\begin{eqnarray}
\hspace{-0.5in} 2\h & \hspace{-0.3in}= & \!\!\!\!
	i \delta E(\h,p,k) \F(\h) + \frac{g^2}{2} \!\! 
	\int \frac{d^2 \q_\perp}{(2\pi)^2}
C(\q_\perp)\Big\{ (2C_s-C_{\rm A})[\F(\h)-\F(\h{-}k\,\q_\perp)] 
        \nonumber \\ && \hspace{0.2in}
        + C_{\rm A}[\F(\h)-\F(\h{+}p\,\q_\perp)] 
        +C_{\rm A}[\F(\h)-\F(\h{-}(p{-}k)\,\q_\perp)] \Big\} , 
\label{eq:integral_eq1}
        \\
\hspace{-1in} \delta E(\h,p,k) &  \hspace{-0.3in}= &
        \frac{\h^2}{2pk(p{-}k)} { + \frac{m_k^2}{2k} +
        \frac{m_{p{-}k}^2}{2(p{-}k)} - \frac{m_p^2}{2p}} \, , \\
\hspace{-0.7in}
C(\q_\perp) &  \hspace{-0.3in}= & 
	\frac{\mD^2}{ \q_\perp^2(\q_\perp^2{+}\mD^2)} \, ,
\quad
\mD^2 = \frac{\gs^2 T^2}{6} (2 N_{\rm c} {+} N_{\rm f}) \, .
\end{eqnarray}
Here $C_{\rm A}=3$ is the adjoint Casimir.  The gluon ``mass'' is the
dispersion correction for a hard gluon, which is $\mD^2/2$.
The production of photons
follows the same equation, but with $C_s \to Q_q^2$, $C_{\rm A} \to
1$, $m_\gamma^2 = 0$, and $g^2 \to e^2$.  These equations then form the
scattering kernel for a system of Boltzmann equations describing the
hard quark and gluon populations in the plasma \cite{Jeon}.

We are in the process of evolving these Boltzmann equations through the
hydrodynamically evolving QGP and applying fragmentation functions at
the surface of hadronization, to evaluate both the jet quenching and the
photon production from the hard secondary partons within a unified
framework.  Our preliminary results indicate that the production rate of
photons from this source will be small, probably negligible compared to
the prompt photons.

\section{Conclusions}

Photons constitute an interesting probe of the Quark Gluon Plasma.
Unfortunately, it may be rather difficult to use this probe, because
many more photons are expected to be produced from the decay of mesons
produced by the fireball, than are produced in the fireball itself.
This is an experimental question.  The theoretical questions are, to
compute the spectrum of produced photons from a piece of QGP, and to
fold this over the hydrodynamical evolution (if indeed this is the
correct description of the plasma's evolution) of the plasma produced in
a heavy ion collision.

The problem of computing the photon production from a glob of QGP turns
out to be more difficult than expected, because collinearly enhanced
bremsstrahlung is a leading order process, and it receives a rather
complicated partial coherence correction called the LPM correction.
This problem is solvable and has now been solved.  The tools used are
similar to what is required to determine the jet quenching rate, and a
unified approach to the two problems is now under way.

\section*{Acknowledgements}

I would like to thank Sangyong Jeon, Peter Arnold, and Larry Yaffe for
collaboration on problems discussed in this talk, and the organizers of
Quark Matter 2004 for inviting me and for arranging a remarkably well
organized and productive meeting.  My work has been supported in part by 
the Natural Sciences and Engineering Research Council of Canada and by
le Fonds Nature et Technologies du Qu\'ebec.

\end{document}